\documentclass[journal]{IEEEtran}
\usepackage{amsmath,amsfonts,mathtools, amsthm, amssymb}
\usepackage{algorithmic}
\usepackage{algorithm}
\usepackage{array}
\usepackage[caption=false,font=normalsize,labelfont=sf,textfont=sf]{subfig}
\usepackage{textcomp}
\usepackage{stfloats}
\usepackage{url}
\usepackage{verbatim}
\usepackage{graphicx}
\usepackage{cite}
\usepackage{multirow}
\usepackage{booktabs}
\newtheorem{lemma}{Lemma}

\newtheorem*{remark}{Remark}
\newtheorem{assumption}{Assumption}
\newcommand{\norm}[1]{\left\lVert #1 \right\rVert}

\newcommand{\supp}[1]{\operatorname{supp}(#1)}

\newcommand{\mstd}[2]{#1\textsubscript{$\pm$#2}}

\begin{document}

\title{A Multi-Scale Attention-Based Attack Diagnosis Mechanism for Parallel Cyber-Physical Attacks in Power Grids}

\author{Junhao Ren, ~\IEEEmembership{Student Member,~IEEE,} Kai Zhao,
    Guangxiao Zhang, ~\IEEEmembership{Member,~IEEE,} Xinghua Liu, ~\IEEEmembership{Senior Member,~IEEE,}
    Chao Zhai, ~\IEEEmembership{Senior Member,~IEEE,} Gaoxi Xiao, ~\IEEEmembership{Senior Member,~IEEE,}
    \thanks{This work was partially supported by the National Research Foundation of Singapore (NRF) through the Future Resilient Systems (FRS-II) Project at the Singapore-ETH Centre (SEC), and partially supported by Ministry of Education, Singapore,
    under contract RG10/23.}
    \thanks{Junhao Ren, Kai Zhao and Gaoxi Xiao are with the School of Electrical
    and Electronic Engineering, Nanyang Technological University, Singapore 639798.
    (E-mail: junhao002@e.ntu.edu.sg, kai007@e.ntu.edu.sg, and egxxiao@ntu.edu.sg.)}
    \thanks{Guangxiao Zhang is with the Institute of Catastrophe Risk Management,
    Nanyang Technological University, Singapore 639798, and also with Future Resilient
    Systems, Singapore-ETH Centre, Singapore 138602. (E-mail: guang xiao.zhang@ntu.edu.sg.)}
    \thanks{Xinghua Liu is with the School of Electrical Engineering, Xi’an University
    of Technology, Xi’an 710048, China. (E-mail: liuxh@xaut.edu.cn.)}
    \thanks{Chao Zhai is with the School of Automation, China University of Geosciences
    (Wuhan), Wuhan 430074, China, and also with the Hubei Key Laboratory of
    Advanced Control and Intelligent Automation for Complex Systems and the
    Engineering Research Center of Intelligent Technology for Geo-Exploration, Ministry
    of Education, Wuhan 430074, China. (E-mail: zhaichao@amss.ac.cn.)}
    \thanks{(\textit{Corresponding author: Gaoxi Xiao.})}
    }

\markboth{Journal of \LaTeX\ Class Files,~Vol.~14, No.~8, January~2026}%
{Shell \MakeLowercase{\textit{et al.}}: A Sample Article Using IEEEtran.cls for IEEE Journals}


\maketitle

\begin{abstract}
  Parallel cyber--physical attacks (PCPA) can simultaneously damage physical transmission lines and disrupt measurement data transmission in power grids, severely impairing system situational awareness and attack diagnosis. This paper investigates the attack diagnosis problem for linearized AC/DC power flow models under PCPA, where physical attacks include not only line disconnections but also admittance modifications, such as those caused by compromised distributed flexible AC transmission system (D-FACTS) devices. To address this challenge, we propose a learning-assisted attack diagnosis framework based on meta--mixed-integer programming (MMIP), which integrates a convolutional graph cross-attention attack localization (CGCA-AL) model. First, sufficient conditions for measurement reconstruction are derived, enabling the recovery of unknown measurements in attacked areas using available measurements and network topology information. Based on these conditions, the attack diagnosis problem is formulated as an MMIP model. The proposed CGCA-AL employs a multi-scale attention mechanism to predict a probability distribution over potential physical attack locations, which is incorporated into the MMIP as informative objective coefficients. By solving the resulting MMIP, both the locations and magnitudes of physical attacks are optimally estimated, and system states are subsequently reconstructed. Simulation results on IEEE 30-bus and IEEE 118-bus test systems demonstrate the effectiveness, robustness, and scalability of the proposed attack diagnosis framework under complex PCPA scenarios.
\end{abstract}


\section{Introduction}
\IEEEPARstart{I}{n} recent years, smart grids have rapidly evolved to enhance the control, monitoring, and efficiency of modern power systems. Smart devices such as Remote Terminal Units (RTUs) and Phasor Measurement Units (PMUs) and modern Supervisory Control and Data Acquisition (SCADA) systems make a major step toward more intelligent and efficient power management. However, this integration of cyber and physical infrastructures also introduces new vulnerabilities. Among them, cyber attacks such as denial-of-service (DoS) \cite{wang2017strategic} and false data injection (FDI) attacks \cite{liang2016review, deng2015defending, zheng2025resilient} pose significant threats by disrupting situational awareness or deceiving control centers.
    
    Beyond such attacks, more sophisticated cyber-physical attacks—simultaneously targeting both cyber and physical layers—can render grid operators’ knowledge obsolete and disable sensing capabilities, preventing timely countermeasures \cite{lee2019Analysis, en15249275}.

    To date, studies in this field have progressed in two main directions. One type
    of cyber-physical attacks is known as coordinated cyber-physical attacks (CCPA)
    \cite{deng2017ccpa}. In CCPA, physical attacks aim to change the topological
    structure of the graph by destroying generators, transmission lines, or transformers,
    while cyber attacks, typically FDI attacks, are launched simultaneously to
    hide physical attack actions until leading to severe devastation
    \cite{chen2010exploring}. In \cite{deng2017ccpa}, defense solutions against
    CCPA were presented using measurements from a limited number of secured
    meters and online tracking of the equivalent impedance. To detect CCPA for a
    discrete-time time-varying power grid at a relatively low computation complexity,
    an attack detector was designed based on exponentially weighted moving
    average statistic, standardization of decision statistics and moving target defense (MTD) \cite{zhou2020real}.
    In \cite{chen2022localization}, Chen et al. proposed a defense algorithm
    which combines MTD with machine learning methods to counter CCPA for a DC
    power flow model while significantly lowering the model dependence. It is noted
    that CCPA is a specific type stealthy attack which requires neat design,
    based on rather comprehensive knowledge of system, to hide the physical attacks
    from measurements for a sufficiently long period of time. There is another
    type of cyber-physical attack, named joint cyber-physical attacks (JCPA),
    proposed in \cite{soltan2015joint}. The significant difference between JCPA and
    CCPA is that, in JCPA, the cyber attacks are to block the measurement data transmission
    from the attacked area rather than hiding the existence of attack, and hence
    may not necessarily request excessive knowledge of system. In \cite{soltan2015joint},
    an LP algorithm was presented to localize the attacks caused by JCPA under
    certain conditions (that the network topology is well-supported after the attack).
    The authors later further extended their results for scenarios involving measurement
    noise and uncertainties \cite{soltan2018power}. For both of these two
    results, it was assumed that no cycle exists within the attacked
    area. Further, an iterative estimation algorithm based on a linear minimum
    mean square error (MMSE) estimation was developed in \cite{hossain2019line}
    to locate failures under JCPA. Compared to other types of attacks, JCPA are
    more direct and malicious while requiring less prior knowledge of systems, which may pose greater pressure on attack localization. Without complete measurements of power injection (in islanding cases), failure localization becomes rather challenging. To address this problem, Huang et al. presented a defense method adopting an LP algorithm and a voting verification algorithm. The voting verification algorithm
    is capable of verifying the attack localization derived from the LP algorithm \cite{huang2022link}.

    To summarize, existing results on scenarios with coexistence of cyber- and physcical-attacks typically assumed that physcical attacks are of link-cut type while  the most damaging physical attacks on a power grid often involve altering line admittances (refer to \cite{zhai2020an}). Therefore, the notion of PCPA is proposed to generalize JCPA from link-cut attacks to more destructive admittance-manipulation cases. This generalization increases the complexity of the attack-diagnosis problem because it drastically expands the feasible solution space of traditional JCPA and CCPA methods, thereby motivating the development of a new, robust diagnosis algorithm tailored to these threats.

    In this study, we present an attack diagnosis framework based on the mutli-scale attention-enhanced MMIP approach for a simplified linearized AC (or DC) power flow model under PCPA. Our main contributions are threefold.
    \begin{itemize}
        \item Compared to existing work \cite{deng2017ccpa, zhou2020real, chen2022localization, soltan2015joint, soltan2018power, hossain2019line, huang2022link}, the proposed PCPA model requires substantially less system knowledge for the adversary and allows non-sparse physical manipulations by modifying line admittances. This leads to more realistic and severe attack scenarios in which the control center completely loses accurate observability over the attacked region.
        
        \item We extend the binary integer programming (BIP) formulation commonly used in JCPA studies \cite{soltan2015joint, huang2022link} to an MMIP model. The MMIP formulation alleviates the difficulties that arise when transitioning from qualitative (physical attack localization) to quantitative (physical attack diagnosis) analysis, and it preserves solution accuracy by introducing weight vectors even when the feasible solution space expands.
        
        \item  We integrate a convolutional graph cross-attention physical attack localization (CGCA-AL) module to improve the efficiency and accuracy of the MMIP formulation. The CGCA-AL algorithm performs the qualitative analysis by estimating the probability of physical attacks on each transmission line within the attacked area. This probability vector is incorporated as a weight in the MMIP objective function to conduct quantitative analysis, guiding the solver toward more accurate solutions. This hybrid learning–optimization approach not only demonstrates superior capability in addressing complex diagnosis problems compared to traditional optimization algorithms but also provides greater interpretability than purely learning-based methods.
    \end{itemize}

    Table \ref{notations} summarizes various notations and abbreviations. The rest of the paper is organized as follows. Section II discusses basic graph theory related to the power flow model and formulates the problem under PCPA.
    Section III establishes the reconstruction conditions for unknown measurements within the attacked area and analyzes basic idea of problem formulation of MMIP. Section IV describes the MMIP-based attack diagnosis framework, including the CGCA-AL algorithm and the objective design of MMIP. In Section V, we validate the effectiveness of the attack diagnosis framework through numerical experiments. Finally, Section VI concludes this work and points
    out some future research directions.

\begin{table}
\caption{Notations and Abbreviations}
\label{notations}
\setlength{\tabcolsep}{3pt}
\begin{tabular}{|p{40pt}|p{200pt}|}
\hline
Notation & Description \\
\hline
$\mathbb{R}$  & The space of real numbers \\
$\vec{(\cdot)}$ & The vector while $\cdot$ represents any lowercase letters \\
$\supp{\cdot}$ & The set of indices of non-zero entries in a vector \\
$[\vec{\cdot}]$ & The diagonalization of vector $\vec{\cdot}$ \\
$| \cdot |$ & The number of the elements in set $\cdot$ \\
$\Gamma$ & The diagonal matrix $[\tfrac{1}{r_{uv}}]_{\{u, v\} \in \mathcal{E}}$ ($r_{uv}$: reactance of line $\{u, v\}$) \\
$I$ & The identity matrix with compatible dimension \\
$\odot$ & The Hadamard product \\
$\|$ & The concatenation operation \\
$\mathbf{0}$ or $\mathbf{1}$ & The vector with all elements $0$ or $1$ \\
$\|\vec{x}\|_p$ & The $\ell_p$-norm of the vector $\vec{x} \in \mathbb{R}^n$ \\
$\|\vec{x}\|_{1, \vec{w}}$ & The weighted $\ell_1$-norm of the vector $\vec{x}$ for $\vec{x}, \vec{w} \in \mathbb{R}^n$, and $\|\vec{x}\|_{1, \vec{w}} \triangleq \sum_{i=1}^{n}{w_i|x_i|}$ \\
\hline
\end{tabular}
\end{table}

\section{Problem Formulation}
    \subsection{Power grid model}
    In this paper, we investigate a linearized (or DC) power flow model which is
    a simplified form of the non-linear AC power flow model. The power grid is characterized
    by the undircted graph $\cal{G}= \left(V, E \right)$, where
    ${\cal V}=\{1, 2, 3,..., n\}$ and ${\cal E}=\{e_{1}, e_{2}, ..., e_{m}\}$
    are the sets of nodes and edges corresponding to the buses and transmission
    lines, respectively. For each transmission line $e_{i}$ connecting a pair of
    buses $\{ u, v\}$, the reactance of $e_{i}$ is denoted as $r_{uv}$ (the same
    as $r_{vu}$). By arbitrarily designating a direction for each line\footnote{It
    is noted that the designated direction will not affect the effectiveness of
    the presented method in this paper.}, the topology of $\cal G$ can be
    represented by the incidence matrix
    ${\cal D}\in \{-1, 0, 1\}^{|{\cal V}| \times |{\cal E}|}$, of which the $(i,
    j)$-th element is defined as:
    \begin{equation}
        {\cal D}_{ij}=
        \begin{dcases}
            1  & \text{if line $e_{j}$ starts from bus $v_{i}$}, \\
            -1 & \text{if line $e_{j}$ ends at bus $v_{i}$},     \\
            0  & \text{otherwise}.
        \end{dcases}
    \end{equation}

    According to \cite{soltan2015joint}, it is known that, given the power supply/demand
    vector $\vec{p}\in \mathbb{R}^{|{\cal V}|}$ and the reactance values, a
    power flow is a solution $P \in \mathbb{R}^{|{\cal V}| \times |{\cal V}|}$
    and $\vec{\theta}\in \mathbb{R}^{|{\cal V}|}$ of :
    \begin{subequations}
        \begin{align}
             & \sum_{v \in {\cal N}_u}{p_{uv}}=p_{u}, \forall u \in{\cal V}\label{eq:power_balance}        \\
             & {\theta}_{u}-{\theta}_{v}-r_{uv}p_{uv}=0, \forall \{u,v\} \in{\cal E}\label{eq:phase_angle}
        \end{align}
    \end{subequations}
    where ${\cal N}_{u}$ is the set of neighbors of bus $u$, $p_{uv}$ is the power
    flow from bus $u$ to bus $v$, and $\theta_{u}$ is the voltage phase angles
    of bus $u$. In the calculation of power flow, the uniqueness of the solution
    for \eqref{eq:power_balance} and \eqref{eq:phase_angle} is guaranteed if
    the demand and the supply are balanced for each bus in a connected graph.
    Hence, within the entire power network, the relationship between the voltage
    phase angles $\vec{\theta}$ and the active power injection $\vec{p}$ is
    delineated as follows:
    \begin{equation}
        A \vec{\theta}= \vec{p}
    \end{equation}
    where $A \in \mathbb{R}^{|{\cal V}| \times |{\cal V|}}$ is the admittance matrix
    of $\cal G$, defined as:
    \begin{equation}
        A_{uv}=
        \begin{dcases}
            0,                                & \text{if}\ u \neq v \ \text{and}\ \{u, v\} \notin{\cal E}, \\
            -1/r_{uv},                        & \text{if}\ u \neq v \ \text{and}\ \{u, v\} \in{\cal E},    \\
            -\sum_{w \in {\cal N}_u}{A_{uw}}, & \text{if}\
 u = v.
        \end{dcases}
    \end{equation}
    In this paper, the SCADA system of the power grid,, as described in \cite{huang2022link}, can
    be simplified as the control center.
    \begin{figure}
        \centering
        \includegraphics[width=0.9\columnwidth]{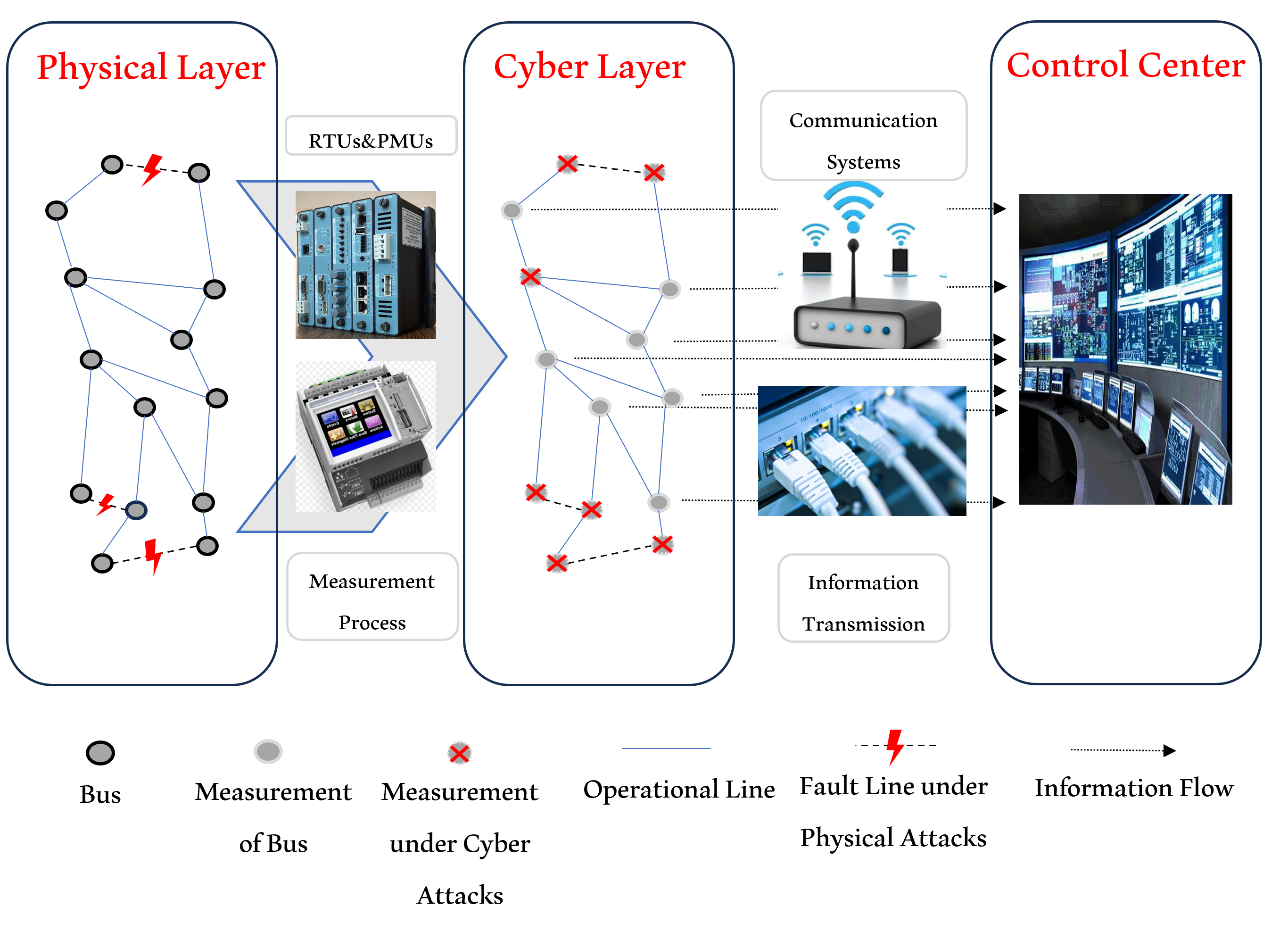}
        \caption{Architecture of power grid and attack scheme of PCPA.}
        \label{fig:wholesystem}
    \end{figure}

    For sensor equipment, we assume the existence of a sensing configuration that guarantees global observability of the power network under normal operating conditions. Under this configuration, the control center can access voltage phase angles and power injection information for all buses, either directly or through state estimation. All available measurements are transmitted to the control center through the cyber communication layer.

    Under cyber attacks, we assume that measurements associated with buses not compromised by the attack remain accurate and available, while measurements from attacked buses may become unavailable or unreliable. This assumption allows the control center to monitor system states and make operational decisions based on the remaining trustworthy measurements.

    \subsection{Attack model of PCPA}
    In this study, a model of simultaneous attacks on both the cyber and physical
    layers is considered. Specifically, on the physical layer, the adversary's
    goal is to compromise certain transmission lines in the targeted area of the
    power grid, resulting in the admittance altering or line disconnections. On
    the cyber layer, the adversary employs various methods, including severing
    specific communications or executing DoS attacks on servers, to obstruct the
    transmission of measurement data from RTUs and PMUs within the attacked area
    to the control center. A successful PCPA can disrupt the stable operation of
    the power grid and hinder the localization of physical attacks. It is essential
    to highlight that PCPA can alter power injections on buses due to the islanding
    effect. The difficulty of attack diagnosis for PCPA lies in the diversity of
    attack mechanisms, which are not restricted to link cut. This expansion of
    attacking capabilities, combined with cyber (DoS) attacks, threaten to
    drastically weaken the resilience of power grids. PCPA poses significant challenges
    to system monitoring and stable operation. It necessitates the control
    center the capabilities of robust failure localization and state
    identification, especially when measurements are partially compromised.

    The attack mechanism of PCPA is shown in Fig. \ref{fig:wholesystem}. It is assumed that after an attack the entire power gird still operates steadily without disastrous cascading buses collapse and the self-protection function has been activated on the attacked buses and transmission lines. Without loss of generality, we cluster all nodes and transmission lines affected by physical attacks and cyber attacks as the attacked area.
    The attacked area is indicated by $H$ and the rest healthy part of the power grid by $\bar{H}={\cal G}{\setminus}H$. It is assumed that
    ${\cal V}_{H}= \{1, 2, 3,..., |{\cal V}|_{H}\}$ and
    ${\cal E}_{H}=\{e_{1}, e_{2}, ..., e_{|{\cal E}_H|}\}$. Therefore, for the
    graph $\cal G$ and its subgraphs $H$ and $\bar H$, $A_{H| \cal G}$ denotes the
    submatrix of$A$ with rows from ${\cal V}_{H}$, and $A_{\bar H | \bar H}$ denotes
    the submatrix of $A$ with rows from ${\cal V}_{H}$ and columns from ${\cal V}
    _{\bar H}$. For instance,
    \[
        A =
        \begin{bmatrix}
            A_{H| \cal{G}}        \\
            A_{\bar{H} | \cal{G}}
        \end{bmatrix}
        =
        \begin{bmatrix}
            A_{H| H}       & A_{H| \bar{H}}      \\
            A_{\bar{H}| H} & A_{\bar{H}|\bar{H}}
        \end{bmatrix}.
    \]
    The values representing post-attack measurements are denoted by a prime $(')$.
    For example, $\vec{\theta}'$ represents the voltage phase angles of the post-attack
    buses.

    In the next section, we provide the theoretical analysis to guide the design
    of attack localization algorithm under PCPA.

    \section{State Analysis of Power Grids under PCPA}
    In this section, the internal situation of power grids under PCPA is analyzed from the
    defense perspective. Specifically, we analyze the state of the system within the $H$
    area based on the information observed in the $\bar H$ area, which provides theoretical guidelines for designing the attack localization algorithm.

    In this work, the change in supply/demand is taken into account (as some buses
    may be disconnected after attacks). Define $\vec{\Delta}= \vec{p}-\vec{p}'$ as
    the supply/demand changes resulted from the physical attacks and $F= \{e_{i}\}
    \subseteq{\cal E}_{H}$ as the set of attacked transmission lines. It is
    known that $A ={\cal D}\Gamma{\cal D}^{T}$ and the post-attack admittance is
    therefore derived as $A' = A -{\cal D}\Gamma [\vec{x}]{\cal D}^{T}$, where $\vec
    {x}$ is an attacking vector such that $\vec{x}_{e}\in (0,1]$ if and only if $e
    \in F$. Therefore, we obtain
    \begin{equation}
        \label{addimitancematrixchange}
        \begin{aligned}
            \vec{\Delta}= & A \vec{\theta}- A' \vec{\theta'}                                                  \\
            =             & A(\vec{\theta}- \vec{\theta'}) +{\cal D}\Gamma[\vec{x}]{\cal D}^{T}\vec{\theta}'.
        \end{aligned}
    \end{equation}
    Apparently, $\vec{\Delta}$ can be partitioned into $\vec{\Delta}_{\bar H}$
    and $\vec{\Delta}_{H}$. Therefore, for the attacked area $H$, we obtain $\vec
    {\Delta}_{H}$ according to \eqref{addimitancematrixchange}:
    \begin{equation}
        \label{eq:6}
        \begin{aligned}
            \begin{bmatrix}\vec{\Delta}_{H}\\ \vec{\Delta}_{\bar H}\end{bmatrix} = & \begin{bmatrix}A_{H|\cal{G}}\\ A_{\bar H |\cal{G}}\end{bmatrix}(\vec{\theta} - \vec{\theta'}) + \begin{bmatrix}D'_{H|\cal{G}}\\ D'_{\bar H|\cal{G}}\end{bmatrix} \vec{x} \\
            \begin{bmatrix}\vec{\Delta}_{H}\\ \vec{\Delta}_{\bar H}\end{bmatrix} = & \begin{bmatrix}A_{H| \cal{G}}\\ A_{\bar H |\cal{G}}\end{bmatrix}(\vec{\theta} - \vec{\theta'}) + \begin{bmatrix}D'_{H}\vec{x}_{H}\\ \textbf{0}\end{bmatrix},
        \end{aligned}
    \end{equation}
    where ${\cal D}' ={\cal D}\Gamma [{\cal D}^{T}\vec \theta']$ and
    ${\cal D}'_{H}$ represents the submatrix of $\cal{D}'$ with rows from ${\cal V}_{H}$ and columns from ${\cal E}_{H}$. 
    
    \subsection{Data reconstruction for the attacked area}
    In this part, our objective is to explore under what conditions missing measurements within the attacked area can be reconstructed using observed information and known topology, thus facilitating accurate attack localization for transmission lines.
    \subsubsection{unknown phase voltage angle $\vec{\theta}'_H$}
    Before further discussion, let us introduce the assumption that restrict the range of the attacked area, thereby ensuring that the problem is solvable.
    \begin{assumption}
        \label{assum:1} Assume that $|{\cal V}_{H}| \le |{\cal V}_{\bar H}|$.
    \end{assumption}
    
    Similarly to the work of Soltan et al. \cite{soltan2015joint}, we have the following lemma based on the $\bar H$ part of \eqref{eq:6}.
    \begin{lemma}
        \label{IdentifyFBus} $\operatorname{supp}(A(\vec{\theta}- \vec{\theta}')-
        \vec{\Delta}) \subseteq{\cal V}_{H}$.
    \end{lemma}
    According to Lemma \ref{IdentifyFBus}, we further give a sufficient reconstruction condition for  the phase angles of the buses in the attacked area when the physical attacks of PCPA are more general than JCPA and CCPA.
    \begin{lemma}
    \label{le: 2}
        Suppose that Assumption \ref{assum:1} holds. The phase angles
        $\vec{\theta}'$ within the attacked area can be correctly reconstructed if
        $A_{\bar{H}|H}$ has a full column rank.
    \end{lemma}

    This sufficient condition is related to the topology structure of the power grid. According to [Corollary 2, \cite{soltan2015joint}], we know that if there is
    a matching in ${\cal G}[{\cal V}_{H},{\cal V}_{\bar H}]$ that covers ${\cal V}
    _{H}$, then $A_{\bar{H}|H}$ has almost surely a full column rank. That is, $\vec{\theta}'$ can be reconstructed almost surely. Therefore, we make the following assumption to reconstruct $\vec
        {\theta}_{H}'$.
    \begin{assumption}
        \label{assum:2} There is a matching in ${\cal G}\left[{\cal V}_{H},{\cal V}
        _{\bar H}\right]$ that covers ${\cal V}_{H}$.
    \end{assumption}
    It is worth noting that this assumption serves as a relatively relaxed sufficient
    prerequisite for the reconstruction of voltage phase angles, and it only falls
    short in certain extreme scenarios (see Appendix~\ref{appen:assump2}). In this study, we only consider those cases where Assumption \ref{assum:2} holds.
    
    \subsubsection{unknown power injection $\vec{p}'_H$}
    It is known that the existence of islanding can be ruled out if the bus power
    injection $\vec{p}_{\bar H}$ within the $\bar H$ area remains unchanged before and after the attack, and vice versa. For instance, if $\vec{p}'_{\bar H}$ differs from $\vec{p}_{\bar H}$ ($\vec{\Delta}_{\bar H}\neq \vec{0}$), we conclude that islanding exists within the $H$ area;
    otherwise, no islanding is present. Such an assessment approach shall allow
    us to determine the status within the attacked area for both islanding and
    islanding-free scenarios.

    For possible islanding cases, we should propose some constraints related to changes
    in power supply/demand after attacks, in accordance with the settings
    established in \cite{huang2022link} and \cite{kundur1994power}.
    \begin{assumption}
        \label{assump:3} The changes on the power supply/demand of the buses are
        balanced and satisfy:
        \begin{align}
             & p_{v}\le p_{v}' \le 0, \forall v \in \{ u| u \in{\cal V}_{H}, p_{u}\le 0\}, \label{eq:assum31} \\
             & p_{v}\ge p_{v}' \ge 0, \forall v \in \{ u| u \in{\cal V}_{H}, p_{u}\ge 0\}, \label{eq:assum32} \\
             & {p}'_{u}/p_{u}={{p}}'_{v}/{p}_{v}, \forall u \in {\cal V}_{H} \ \text{and} \ \forall v \in {\cal V}_{H}, \label{eq: powerrecovery} \\
             & \textbf{1}^{T}\vec \Delta = 0. \label{eq:assum33}
        \end{align}
    \end{assumption}
    In this assumption, it guarantees the same bus type and balanced power injection before and after PCPA. In addition, \eqref{eq: powerrecovery} represents the \textit{proportional load shedding/generation reduction} assumption (see \cite{kundur1994power}), which has been widely adopted in studies of small- and medium-scale power grids \cite{girgis2010application}.

    Based on Assumption \ref{assump:3}, we present the following lemma to reconstruct $\vec{p}'_H$ within the $H$ area.
    \begin{lemma}
        \label{le: deltarecovery} Suppose Assumptions \ref{assum:2} and \ref{assump:3}
        hold. For all $u \in{\cal V}_{H}$, there exists $v \in{\cal V}_{\bar H}$
        such that $e = (u, v) \in \cal{E}$. The post-attack power injection
        $p'_{u}$ can be recovered according to \eqref{eq: powerrecovery}.
    \end{lemma}

    It is noted that, in this paper, reconstructing $\vec{\theta}'_H$ is necessary; without it, the problem may be unsolvable. In contrast, reconstructing of $\vec{p}'_{H}$ is optional, although it can enhance the efficiency of the proposed attack diagnosis framework. Further details are provided in the next section.

    \subsection{Attack Diagnosis Formulation for PCPA}
    Once the critical unknown measurements ($\vec{\theta}'_H$) are reconstructed, we may proceed to tackle the attack localization problem for PCPA. Generally, to locate the set of attacked lines $F$, we need to solve a complex optimization problem with a series of linear matrix inequalities (LMI) constraints. Conventionally, for cases involving only link disconnections (JCPA), the problem can be formulated as a binary linear programming problem as follows \cite{huang2022link}:
    \begin{align}
        \text{(P0):}\quad & \min_{\vec{x}_H, \vec{\Delta}_H}\norm{\vec{x}_H}_{1}                     \\
        \text{s.t.}\quad  & \eqref{eq:6}, \eqref{eq:assum31}, \eqref{eq:assum32}, \eqref{eq: powerrecovery}, \eqref{eq:assum33} \\
                          & x_{e}\in \{0, 1\}, \forall e \in{\cal E}_{H}
    \end{align}
    The optimization problem (P0) is NP-hard and is formulated based on a \textit{sparsity} assumption. When $x_e \in \{0,1\}$, the $\ell_1$ objective is equivalent to minimizing the cardinality of the faulty-link set, which encourages the solution $\bar{\vec{x}}_H^\ast$ to align with the true attack pattern $\vec{x}_H^\ast$.

    In this paper, more general physical attacks, including general admittance alteration and line disconnection as its special case, are considered. This generalization induces several fundamental changes to problem (P0): (i) \textit{the feasible domain is relaxed from the discrete set $\{0,1\}$ to the continuous interval $[0,1]$;} (ii) \textit{sparsity in the strict sense of the number of faulty edges is no longer preserved, since the $\ell_1$ objective now measures the aggregate impairment level of the attacked-link set;} and (iii) \textit{as a consequence, multiple distinct impairment patterns may fit the measurements with comparable objective values, leading to ambiguity and ill-posedness in attack localization.} Although this relaxation improves numerical tractability by enabling continuous optimization, it necessitates different additional mechanisms to guide the solution toward physically meaningful attack patterns.

    To address this issue, prior information is incorporated into the objective function. By contrast to (P0), the prior-guided formulation (P1) in the form of MMIP reshapes the solution space and steers the optimizer toward physically plausible attack patterns under the general attack model.
    \begin{align}
        \text{(P1):}\quad & \min_{\vec{x}_H, \vec{\Delta}_H}\|\vec{x}_{H}\|_{1, \vec{c}}  \label{eq:p1}       \\
        \text{s.t.}\quad  & \eqref{eq:6}, \eqref{eq:assum31}, \eqref{eq:assum32}, \eqref{eq: powerrecovery}, \eqref{eq:assum33}, \\
                          & 0 \le \vec{x}_{H}\le 1,
    \end{align}
    where $\vec{c}$ is a weight vector determined by informative priors obtained from the power grid state, and $0 \le \vec{x}_{H} \le 1$ denotes element-wise inequalities.

    Essentially, the objective of (P1) is to penalize aggregate line impairment while favoring solutions that are consistent with the prior likelihood of attacks on different transmission lines. By assigning different weights in $\vec{c}$, the optimization is guided to concentrate impairment on a small set of high-risk lines, thereby mitigating the ambiguity introduced by the continuous relaxation. The accuracy of $\vec{c}$ directly affects the reliability of the solution obtained from (P1). The construction of $\vec{c}$ from observed measurements is described in the next section.

\section{Algorithm Design for Attack Diagnosis}
    In this section, we propose the general attack diagnosis framework to cope with PCPA based on the above discussion.
    \begin{figure}[!t]
        \centering
        \includegraphics[width=0.8\columnwidth, height=7cm]{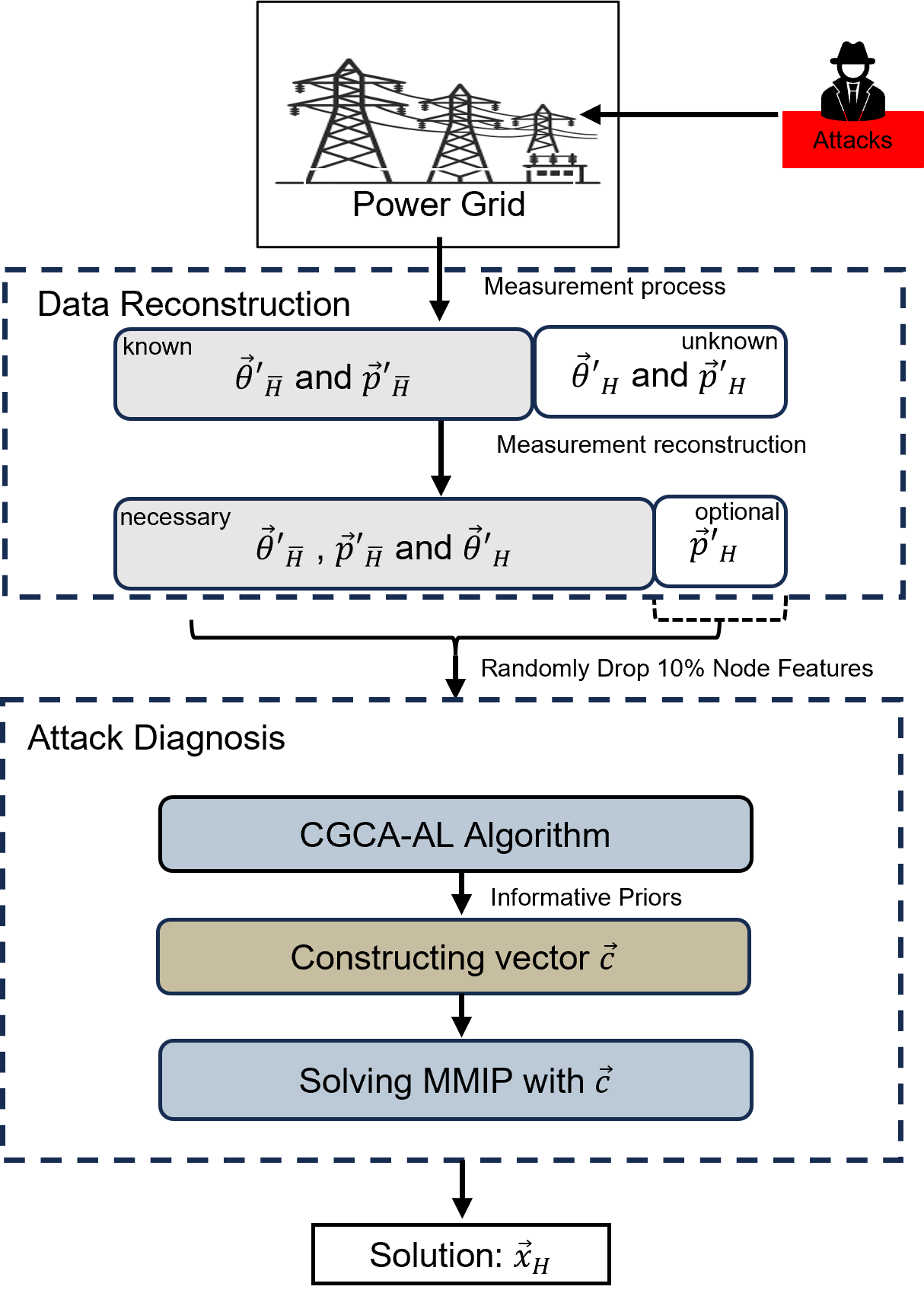}
        \caption{Scheme of Attack diagnosis process for PCPA.}
        \label{fig:1}
    \end{figure}
    In Fig.~\ref{fig:1}, there are three main steps in the proposed attack diagnosis framework: attack localization algorithm, constructing the weight vector $\vec{c}$ and solving MMIP with $\vec{c}$.

    In the attack localization stage, we estimate the distribution of attack locations based on the available measurements. Due to the non-Euclidean and topology-dependent nature of power grids, purely model-based localization approaches may suffer from degraded accuracy or lack robustness when measurements are missing, reconstructed, or corrupted by cyber attacks. This motivates the adoption of a learning-based approach that can map observed system states to probabilistic estimates of attack locations.

    Graph neural networks (GNNs) are designed to learn representations from graph-structured data by aggregating neighborhood information through message passing \cite{boyaci2022joint, takiddin2023generalized, chen2020fault}. This capability enables GNNs to capture changes in spatial correlations of system states (e.g., voltage angles and power injections) induced by physical attacks. Since PCPA can induce significant structural changes in network parameters and effective connectivity, a GAT is adopted to learn how physical attacks on transmission lines within the attacked area $H$ affect spatial correlations of neighbors.

    In addition, it is noted that an attack on a single transmission line may potentially influence the operating states of a large portion of the network due to the strong physical coupling of power grids. Solely capturing local spatial correlations may overlooks the potential significant dependency among long-range nodes, thereby limiting the representation capability of the algorithm. Therefore, we introduce a CNN layer to complement the GAT by capturing long-range dependencies that may not be efficiently modeled through local message passing alone. 

    To further enhance representation learning, we incorporate a multi-head cross-attention (MHCA) module that allows the target node subset ${\cal V}_H$ to query the entire graph. The cross-attention dynamically fuses long-range contextual signals from all nodes into the target subset, emphasizing the most attack-relevant regions of the network. This global, query-driven aggregation complements local spatial modeling and sequence-based dependencies, enabling the model to better capture how physical perturbations propagate through the grid and affect critical nodes.

    \subsection{GAT Module}
    A single GAT layer contains two critical components:
    \subsubsection{Graph attention mechanism}
    For bus $i$ and bus $j$, the observed features are $\vec{h}_{i}\in{\mathbb{R}^{f}}$
    and $\vec{h}_{i}\in{\mathbb{R}^{f}}$, respectively. The attention coefficients
    $\omega_{ij}$ are computed as:
    \begin{equation}
        \label{eq:attentioncoe}\omega_{ij}= \operatorname{LeakyReLU}\left( \vec{z}
        ^{\text{T}}[\boldsymbol{W}\vec{h}_{i}\| \boldsymbol{W}\vec{h}_{j}] \right),
    \end{equation}
    where $\boldsymbol{W}\in{\mathbb{R}}^{f' \times f}$ denotes the shared
    transformation matrix for the feature embedding, a shared attention mechanism
    is parametrized by the weight vector $\vec{z}\in{\mathbb{R}}^{2f'}$, and
    $\operatorname{LeakyReLU}$ is the nonlinear activation function. The attention
    coefficients are then normalized by a softmax function to get the attention scores,
    as shown in \eqref{eq:normalizeattention}.
    \begin{equation}
        \label{eq:normalizeattention}\alpha_{ij}= \frac{\operatorname{exp}(\omega_{ij})}{\sum_{k
        \in {\cal N}_i}{\operatorname{exp}(\omega_{ik})}}.
    \end{equation}
    \subsubsection{Message passing}
    After obtaining the attention scores, message passing is performed to update the feature for each bus by GCN. The output of the GCN at bus $i$ can be expressed as:
    \begin{equation}
        \label{eq: GCN}\vec{h}'_{i}= \sigma \left( \sum_{j \in {\cal N}_i}{\alpha_{ij} \boldsymbol{W} \vec{h}_j}
        \right),
    \end{equation}
    where $\sigma$ denotes the sigmoid activation function. In addition, for multi-head attention mechanism, all of the outputs for different attentions can be aggregated to obtain the final feature representation.

    \subsection{MHCA Mechanism}
    In the cross-attention mechanism, the input features $F_q$, $F_k$ and $F_v$ are first linearly projected into queries, keys and values:
    \begin{equation}
        Q = F_q W_q,\quad
        K = F_k W_k,\quad
        V = F_v W_v,
    \end{equation}
    where $F_q \in \mathbb{R}^{B \times N_q \times d}$, $F_k, F_v \in \mathbb{R}^{B \times N_k \times d}$ denote the query, key and value feature sequences in a mini-batch, respectively. $d$ is the model dimension, and $W_q, W_k, W_v \in \mathbb{R}^{d \times d}$ are learnable projection matrices.
    In our setting, $F_q$ corresponds to the features of the critical node set ${\cal V}_H$, while $F_k$ and $F_v$ collect the distinct features of all nodes in the graph.

    The cross-attention operator is then defined as
    \begin{equation}
        \operatorname{Catt}(Q, K, V)
        = \operatorname{Softmax}\!\left( \frac{Q K^\top}{\sqrt{d_k}}\right) V,
    \end{equation}
    where $d_k$ denotes the dimensionality of key vectors. For the $m$-th head of the multi-head mechanism, the cross-attention operator is defined as $ \mathrm{head}_m = \mathrm{Catt}(Q^{(m)}, K^{(m)}, V^{(m)})$ with dimension $d_h=d / t$. The outputs of all heads are then concatenated along the feature dimension and linearly projected:
    \begin{equation}
        \mathrm{MultiHeadCatt}(Q, K, V)
        = \bigl[ \mathrm{head}_{1}; \dots; \mathrm{head}_{t } \bigr] W_o,
    \end{equation}
    where $W_o \in \mathbb{R}^{d \times d}$ is a learnable output projection matrix.

    \subsection{Architecture of the CGCA-AL algorithm}
\begin{figure*}
        \centering
        \includegraphics[width=0.75\textwidth, height=4cm]{
            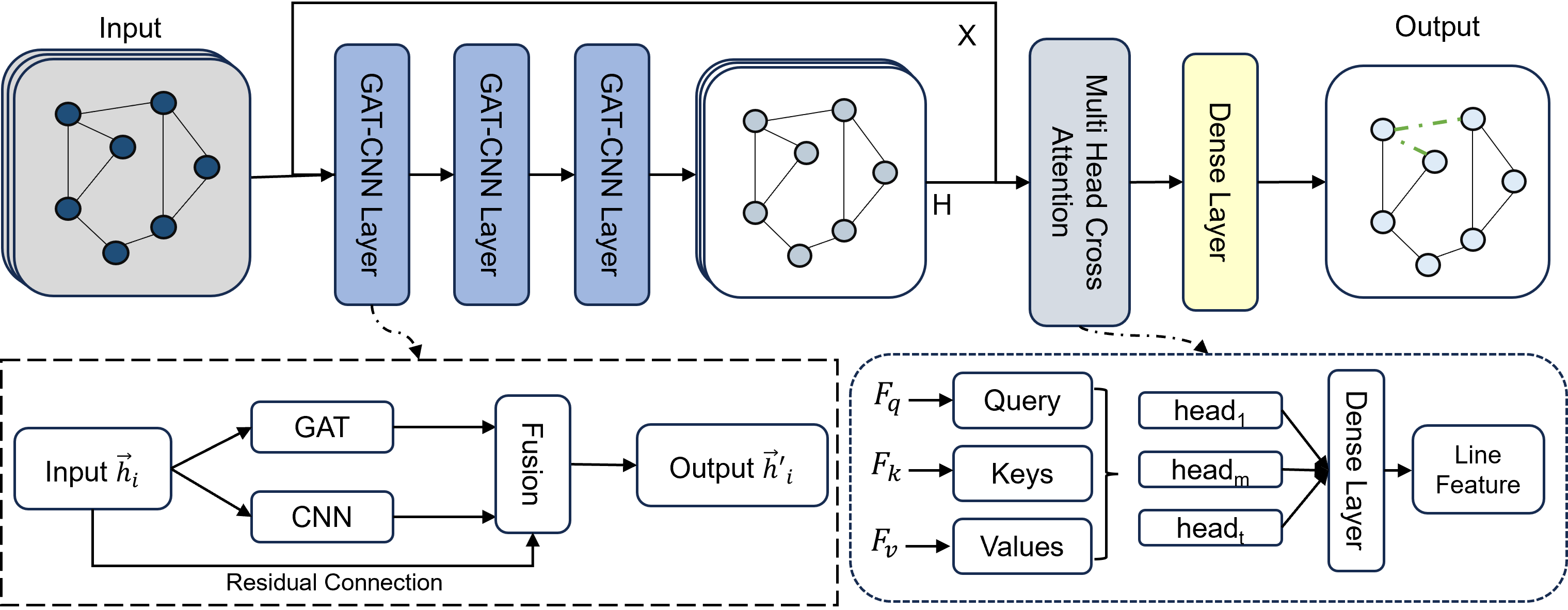
        }
        \caption{The structure of the CGCA-AL algorithm. GAT-CNN layer is used to extract the high dimensional features from local spatial  and long-range correlations, respectively. Multi head cross-attention is utilized to enhanced the high dimensional features of ${\cal V}_H$.}
        \label{fig:faultlocalization}
    \end{figure*}

    The proposed CGCA-AL algorithm aims to localize the attacks caused by PCPA, as shown in Fig.~\ref{fig:faultlocalization}.
    The model ingests the observed bus states of the entire power grid, then stacks $L_{g}$ GAT-CNN layers to jointly  capture local spatial dependencies and long-range patterns, with residual connections in each layer to improve robustness. A subsequent MHCA module refines the high dimensional representations of ${\cal V}_H$ by adaptively aggregating information from ${\cal V}_{\bar H}$. The enhanced node features of ${\cal V}_H$  are then assembled into edge representations for ${\cal E}_H$, which are passed through a dense layer to  estimate the probability $\vec{y}$ that each line in ${\cal E}_H$ is under physcical attack. In the next section, we shall show how to translate $\vec{y}$ into $\vec{c}$ and then identity the actual states of transmission lines within the attacked area with MMIP model given $\vec{c}$.
    
    \subsection{Line State Identification (LSI) Based on MMIP}
    As aforementioned, the islanding caused by physical attacks and the
    existence of the cycles within the $H$ area will definitely bring  challenges, such as the increasing complexity of combinatorial search space, to MMIP. Mathematically,
    once islanding occurs, according to \eqref{eq:6}, we give that
    \begin{equation}
        \label{eq: 20}\vec \Delta_{H}= A_{H|G}(\vec{\theta}- \vec{\theta'}) + D_{H}
        ' \vec{x}_{H}.
    \end{equation}
    After a subtle transformation, \eqref{eq: 20} can be rewritten in terms
    of the pair of $(\vec{x}_{H}, \vec{\Delta}_{H})$.
    \begin{equation}
        \label{eq: 21}\left[I -D_{H}'\right]
        \begin{bmatrix}
            \vec \Delta_{H} \\
            \vec x_{H}
        \end{bmatrix}
        = A_{H|G}(\vec{\theta}- \vec{\theta'}).
    \end{equation}
    \begin{remark}
    \label{re: deltaisoptional}
        Even when $\vec \Delta_{H}$ is available, \eqref{eq: 21} remains underdetermined: the matrix $\left[I -D_{H}'\right]$ can have the rank at most $|{\cal{V}}_{H}|$, which is insufficient to pin down a unique solution. In the more challenging case where $\vec \Delta_{H}$ cannot be reconstructed, the total number of unknowns grows to $|{\cal{V}}_{H}| + |{\cal{E}}_{H}|$, further enlarging the solution space. This inherent underdetermination indicates that missing $\vec \Delta_{H}$ may undermine the efficiency and accuracy of standard MMIP-based LSI. Meanwhile, the necessity of reconstructing $\vec \theta'_{H}$ is clear from \eqref{eq: 21}. Without it, the matrix $\left[I -D_{H}'\right]$ collapses to the zero matrix (rank 0), making it impossible to extract any information related to the attacked area.
    \end{remark}
    In the following, we incorporate the output $\vec{y}$ of CGCA-AL algorithm into the objective function of (P1) so as to steer the solver toward higher‐risk regions. Concretely, by setting $\vec{c}=\textbf{1}-\vec{y}$, we impose larger penalties on lines that the CGCA-AL model predicts as secure, thereby discouraging any candidate solution from grouping these lines as under attacks. The new attack diagnosis problem formulation (P2), in terms of MMIP based on multi-scale attention mechanism (CGCA-AL algorithm), is given by:
    \begin{align}
        \text{(P2):}\quad & \min_{\vec{x}_H, \vec \Delta_{H}}\|\vec{x}_{H}\|_{1, \textbf{1}- \vec{y}}\label{eq:p2}           \\
        \text{s.t.}\quad  & \eqref{eq:assum31}, \eqref{eq:assum32}, \eqref{eq: powerrecovery}, \eqref{eq:assum33}, \eqref{eq: 20}, \\
                          & 0 \le \vec{x}_{H}\le 1,
    \end{align}
    where $0 \le \vec{x}_{H}\le 1$ denotes element-wise inequality. The
    elements of the weight vector $\vec{c}=\textbf{1}-\vec{y}$ represent the probabilities
    that the corresponding transmission lines are free of physical attacks.
    Therefore, the objective function of (P2) can be interpreted as that physical
    attacks more likely have happened on those transmission lines with higher $\vec
    {y}_{i}$ (lower $(1-\vec{y}_{i})$). The complete fault‐diagnosis algorithm is summarized in Algorithm \ref{alg:1}.

    \begin{algorithm}
        \caption{Multi-Scale Attention-Based Attack Diagnosis Framework for Power Grids under PCPA.}
        \label{alg:1}
        \begin{algorithmic}
            [1] \STATE \textbf{Input:} $A$, $D$, $\Gamma$, $\vec{p}$,
            $\vec{p}'_{\bar H}$, $\vec{\theta}$, $\vec{\theta}'_{\bar H}$.
            \STATE \textbf{Output:} Estimated physical attacks $\vec{x}_{H}$.
            \STATE Obtain $\vec{\theta}'_{H}$ $\gets$ Lemma \ref{le: 2}.
            \IF{$\vec{p}_{\bar H}\neq \vec{p}_{\bar H}'$}  
            \STATE  (\textit{Optional}) Obtain $\vec{p}'_{H}$ $\gets$ Lemma \ref{le: deltarecovery}.
            \ELSE \STATE $\vec{\Delta}_{H}\gets \vec{0}$. 
            \ENDIF 
            \STATE Obtain $\vec{y}$ $\gets$ CGCA-AL algorithm
            \STATE Let $\vec{c}=\textbf{1}-\vec{y}$.
            \STATE Obtain $\vec{x}_{H}$ $\gets$ solving MMIP (P2). 
            \RETURN $\vec{x}_{H}$.
        \end{algorithmic}
    \end{algorithm}

    \section{Experimental results}
    In the numerical simulation, we evaluate the proposed attack diagnosis framework from two perspectives: AL and LSI. We carry out simulations of attack diagnosis framework on IEEE 30-bus and IEEE 118-bus cases. The generation process of the data used for training and evaluation is introduced in the following.

    \subsection{Data collection}
    For both IEEE 30- and 118-bus systems, each bus is connected to a load randomly sampled from a load-level distribution derived from a realistic annual residential load profile \cite{muratori2018impact}. To construct PCPA scenarios, the power grid is partitioned into the attacked area $H$ and the remaining area $\bar H$ using a degree-based greedy search (DBGS)  algorithm that satisfies Assumptions~\ref{assum:1} and~\ref{assum:2}.

    For attacked transmission lines, admittance-altering attacks are modeled by multiplying the nominal impedance by a random factor sampled from $(1.5,5)$, while link-cut attacks are approximated by a large impedance scaling factor sampled from $(100,1000)$, effectively emulating line disconnection. All power flow simulations are conducted using MATPOWER \cite{zimmerman2010matpower}.

    For each attack type, $500$ samples are generated by running power flow under randomly sampled load profiles to construct the dataset for training and validation. Each attacked line, whether due to admittance alteration or disconnection, is assigned a label of ``1'', while normal operational lines are labeled ``0'', reflecting the binary nature of attack localization. For testing, $200$ samples are generated for each attack cardinality $|F|$ to evaluate the average performance under different attack intensities. The input features include voltage phase angles $\vec{\theta}$, power injections $\vec{p}$, and loads $\vec{l}$. The graph weight matrix is set as $W=\mathcal A$, where $\mathcal A$ denotes the adjacency matrix of the power grid, ensuring consistency with the underlying network topology.

    \subsection{Numerical results}
    In the following, we illustrate the numerical results obtained by the attack localization algorithm and the LSI algorithm. Details about implementation and metrics please refer to Appendix~\ref{training_details}.

    \subsubsection{The performance on the IEEE 30 buses}
    In this case, we generate a cyclic attacked area $H$ with
    $|{\cal V}_{H}| =8$ and $|{\cal E}_{H}| =8$ using the DBGS algorithm. Table \ref{tab:flcase30} shows the performance of the proposed CGCA-AL algorithm for AL task on IEEE 30-bus cases.
    \begin{table*}[t]
        \centering
        \caption{Performance comparison of AL task for IEEE 30-bus case under different numbers of attacked lines}
        \label{tab:flcase30}
        \renewcommand{\arraystretch}{1.15}
        \setlength{\tabcolsep}{5pt}
        \begin{tabular}{c| c|cccccccc}
        \hline
        \textbf{Algorithm} & \textbf{Metric} & \multicolumn{8}{c}{\textbf{Number of Attacked Lines } $|F|$} \\
         &  & 1 & 2 & 3 & 4 & 5 & 6 & 7 & 8 \\
        \hline
        \multirow{4}{*}{\textbf{GCN}}
        & Accuracy & 0.9194 & 0.8619 & 0.8081 & 0.7825 & 0.7137 & 0.7181 & 0.6550 & 0.6200 \\
        & FAR      & 0.0336 & 0.0350 & 0.0190 & 0.0375 & 0.0650 & 0.0500 & 0.0500 & 0.0000 \\
        & MDR      & 0.4100 & 0.4475 & 0.4800 & 0.3975 & 0.4190 & 0.3592 & 0.3871 & 0.3800 \\
        & $\mathrm{F}_1$    & 0.6466 & 0.6667 & 0.6702 & 0.7348 & 0.7173 & 0.7733 & 0.7566 & 0.7654 \\
        \hline
        \multirow{4}{*}{\textbf{GAT}}
        & Accuracy & 0.9231 & 0.8606 & 0.8075 & 0.7806 & 0.7144 & 0.7219 & 0.6725 & 0.6381 \\
        & FAR      & 0.0264 & 0.0325 & 0.0180 & 0.0437 & 0.0533 & 0.0425 & 0.0550 & 0.0000 \\
        & MDR      & 0.4300 & 0.4600 & 0.4833 & 0.3950 & 0.4250 & 0.3567 & 0.3664 & 0.3619 \\
        & $\mathrm{F}_1$    & 0.6496 & 0.6595 & 0.6681 & 0.7339 & 0.7156 & 0.7763 & 0.7720 & 0.7791 \\
        \hline
        \multirow{4}{*}{\textbf{GAE}}
        & Accuracy & 0.9600 & 0.9131 & 0.8644 & 0.8369 & 0.7963 & 0.7769 & 0.7137 & 0.6900 \\
        & FAR      & 0.0157 & 0.0192 & 0.0160 & 0.0300 & 0.0300 & 0.0350 & 0.0550 & 0.0000 \\
        & MDR      & 0.2100 & 0.2900 & 0.3350 & 0.2963 & 0.3080 & 0.2858 & 0.3193 & 0.3100 \\
        & $\mathrm{F}_1$    & 0.8316 & 0.8034 & 0.7862 & 0.8118 & 0.8094 & 0.8276 & 0.8063 & 0.8166 \\
        \hline
        \multirow{4}{*}{\textbf{Transformer}}
        & Accuracy & 0.9656 & 0.9044 & 0.8575 & 0.8219 & 0.7725 & 0.7338 & 0.6650 & 0.5919 \\
        & FAR      & \textbf{0.0029} & \textbf{0.0117} & \textbf{0.0130} & 0.0300 & \textbf{0.0217} & \textbf{0.0325} & \textbf{0.0200} & 0.0000 \\
        & MDR      & 0.2550 & 0.3475 & 0.3583 & 0.3262 & 0.3510 & 0.3442 & 0.3800 & 0.4081 \\
        & $\mathrm{F}_1$    & 0.8442 & 0.7733 & 0.7715 & 0.7909 & 0.7810 & 0.7870 & 0.7641 & 0.7436 \\
        \hline
        \multirow{4}{*}{\textbf{CGCA-AL}}
        & Accuracy & \textbf{0.9669} & \textbf{0.9156} & \textbf{0.8688} & \textbf{0.8525} & \textbf{0.8181} & \textbf{0.7887} & \textbf{0.7556} & \textbf{0.7375} \\
        & FAR      & 0.0107 & 0.0167 & 0.0180 & \textbf{0.0262} & 0.0317 & 0.0525 & 0.0350 & \textbf{0.0000} \\
        & MDR      & \textbf{0.1900} & \textbf{0.2875} & \textbf{0.3200} & \textbf{0.2687} & \textbf{0.2720} & \textbf{0.2642} & \textbf{0.2743} & \textbf{0.2625} \\
        & $\mathrm{F}_1$    & \textbf{0.8594} & \textbf{0.8085} & \textbf{0.7953} & \textbf{0.8321} & \textbf{0.8334} & \textbf{0.8394} & \textbf{0.8386} & \textbf{0.8489} \\
        \hline
        \end{tabular}
    \end{table*}
    For accuracy and $\text{F}_{1}$ score metrics, CGCA-AL attains the highest values for all attack scenarios. When $|F|=1$, CGCA-AL already achieves an accuracy of $0.9669$ and an $\mathrm{F}_{1}$ score of $0.8594$, outperforming the strongest baseline (Transformer) by approximately $2.3\%$ in $\mathrm{F}_{1}$. As the number of attacked lines increases, this advantage becomes more pronounced. For $|F|\geq 3$, CGCA-AL maintains an $\mathrm{F}_{1}$ score above $0.79$, while other baselines exhibit a faster performance degradation. Notably, when $|F|=8$, CGCA-AL preserves an $\mathrm{F}_{1}$ score of $0.8489$, exceeding GAE by more than $3\%$. For FAR and MDR metrics, CGCA-AL also consistently yields the lowest MDR across all attack scales and reduces missed detections by approximately $6\%$–$13\%$ compared to GAE and Transformer for moderate to large $|F|$. Although the Transformer achieves a slightly lower FAR for small $|F|$, the gap between it and CGCA-AL algorithm is below $2\%$.

    The performance of the attack diagnosis scheme combining the CGCA-AL algorithm with the LSI algorithm is further evaluated, and the results are summarized in Table~\ref{tab:lsicase30}. As observed, \textbf{CGCA-AL+LSI consistently achieves the lowest or near-lowest estimated error across all attack scales}, demonstrating superior attack diagnosis capability. In most cases, it provides a performance improvement ranging from at least $2\%$ up to $39\%$ compared with learning-based baselines. Moreover, CGCA-AL+LSI yields less than half of the estimated error of the model-based FLD method, highlighting its significant advantage in both accuracy and robustness.
    \begin{table*}[t]
    \centering
    \caption{Estimated error of LSI task for IEEE 30-bus case under different numbers of attacked lines}
    \label{tab:lsicase30}
    \scriptsize
    \renewcommand{\arraystretch}{1.05}
    \setlength{\tabcolsep}{3.5pt}
    \resizebox{\textwidth}{!}{%
    \begin{tabular}{c|cccccccc}
    \hline
    \textbf{Algorithm} & \multicolumn{8}{c}{\textbf{Number of Attacked Lines} $|F|$} \\
    & 1 & 2 & 3 & 4 & 5 & 6 & 7 & 8 \\
    \hline
    \textbf{FLD} &
    \mstd{0.0703}{0.2575} & \mstd{0.1064}{0.2441} & \mstd{0.1811}{0.2768} & \mstd{0.2245}{0.2713} &
    \mstd{0.2548}{0.2546} & \mstd{0.2399}{0.2227} & \mstd{0.2890}{0.1960} & \mstd{0.2798}{0.1806} \\

    \textbf{GCN+LSI} &
    \mstd{0.0000}{0.0000} & \mstd{0.0243}{0.1345} & \mstd{0.0509}{0.1792} & \mstd{0.0884}{0.2088} &
    \mstd{0.1220}{0.2223} & \mstd{0.1227}{0.1852} & \mstd{0.1306}{0.1667} & \mstd{0.1681}{0.1597} \\

    \textbf{GAT+LSI} &
    \mstd{0.0000}{0.0000} & \mstd{0.0243}{0.1345} & \mstd{0.0509}{0.1792} & \mstd{0.0883}{0.2088} &
    \mstd{0.1222}{0.2225} & \mstd{0.1222}{0.1855} & \mstd{0.1302}{0.1670} & \mstd{0.1686}{0.1603} \\

    \textbf{GAE+LSI} &
    \mstd{0.0000}{0.0000} & \mstd{0.0199}{0.1217} & \textbf{\mstd{0.0399}{0.1619}} & \mstd{0.0653}{0.1897} &
    \mstd{0.1025}{0.2138} & \mstd{0.1140}{0.1819} & \mstd{0.1156}{0.1565} & \mstd{\textbf{0.1677}}{0.1591} \\

    \textbf{Transformer+LSI} &
    \mstd{0.0000}{0.0000} & \textbf{\mstd{0.0198}{0.1208}} & \mstd{0.0469}{0.1752} & \mstd{0.0746}{0.1982} &
    \mstd{0.1080}{0.2149} & \mstd{0.1214}{0.1852} & \mstd{0.1232}{0.1655} & \mstd{0.1696}{0.1603} \\

    \textbf{CGCA-AL+LSI} &
    \textbf{\mstd{0.0000}{0.0000}} & \mstd{0.0199}{0.1217} & \mstd{0.0414}{0.1679} & \textbf{\mstd{0.0632}{0.1863}} &
    \textbf{\mstd{0.0979}{0.2074}} & \textbf{\mstd{0.1088}{0.1772}} & \textbf{\mstd{0.1134}{0.1556}} & \mstd{0.1680}{0.\textbf{1572}} \\
    \hline
    \end{tabular}%
    }
    \end{table*}

    \subsubsection{The performance on IEEE 118-bus}
    In this case, we generate an attacked area $H$ with $|{\cal V}_{H}| =20$, $|{\cal E}
    _{H}| =24$ and at  least $5$ cycles using the DBGS algorithm. 
    The AL performance of the proposed CGCA-AL algorithm is reported in Table~\ref{tab:flcase118_1_24}. As $|F|$ increases, CGCA-AL exhibits a monotonic improvement in the $\mathrm{F}_1$ score from $0.7654$ to $0.9947$, consistently outperforming all baseline methods and achieving the highest accuracy and $\mathrm{F}_1$ across all attack scales. Quantitatively, under moderate attack scenarios ($|F|=6$--12), CGCA-AL improves the $\mathrm{F}_1$ score by approximately $8\%$--$15\%$ compared with Transformer- and GAT-based baselines, while under dense attack conditions ($|F|\geq 18$), the improvement remains at $5\%$--$10\%$, demonstrating strong scalability. Moreover, CGCA-AL achieves the lowest or near-lowest FAR and MDR in most cases; for example, at $|F|=12$, the MDR is reduced by $10\%$ compared with GCN, highlighting its enhanced robustness against both FAR and MDR.
    \begin{table*}[t]
      \centering
      \caption{Performance comparison of AL task for IEEE 118-bus case under different numbers of attacked lines ($|F|=1$--$24$).}
      \label{tab:flcase118_1_24}
      \footnotesize
      \renewcommand{\arraystretch}{1.12}
      \setlength{\tabcolsep}{4.5pt}
      \resizebox{\textwidth}{!}{%
      \begin{tabular}{c|c|cccccccccccc}
      \hline
      \textbf{Algorithm} & \textbf{Metric} & \multicolumn{12}{c}{\textbf{Number of Attacked Lines} $|F|$} \\
      \hline
      
      & & 1 & 2 & 3 & 4 & 5 & 6 & 7 & 8 & 9 & 10 & 11 & 12 \\
      \hline
      \multirow{4}{*}{\textbf{GCN}} 
      & Accuracy & 0.9788 & 0.9554 & 0.9304 & 0.9077 & 0.8823 & 0.8621 & 0.8396 & 0.8221 & 0.8140 & 0.7790 & 0.7702 & 0.7510 \\
      & FAR      & 0.0017 & 0.0045 & 0.0102 & 0.0165 & 0.0261 & 0.0433 & 0.0529 & 0.0653 & 0.0750 & 0.1361 & 0.1292 & 0.1808 \\
      & MDR      & 0.4700 & 0.4850 & 0.4850 & 0.4713 & 0.4660 & 0.4217 & 0.4214 & 0.4031 & 0.3711 & 0.3400 & 0.3486 & 0.3171 \\
      & $F_1$    & 0.6752 & 0.6581 & 0.6492 & 0.6563 & 0.6540 & 0.6771 & 0.6778 & 0.6910 & 0.7171 & 0.7133 & 0.7221 & 0.7328 \\
      \hline
      
      \multirow{4}{*}{\textbf{GAT}} 
      & Accuracy & 0.9796 & 0.9535 & 0.9275 & 0.9062 & 0.8869 & 0.8615 & 0.8467 & 0.8319 & 0.8240 & 0.7975 & 0.7833 & 0.7781 \\
      & FAR      & 0.0037 & 0.0125 & 0.0214 & 0.0305 & 0.0366 & 0.0531 & 0.0624 & 0.0688 & 0.0780 & 0.1214 & 0.1062 & 0.1321 \\
      & MDR      & 0.4050 & 0.4200 & 0.4300 & 0.4100 & 0.4040 & 0.3950 & 0.3743 & 0.3669 & 0.3394 & 0.3160 & 0.3473 & 0.3117 \\
      & $F_1$    & 0.7083 & 0.6754 & 0.6628 & 0.6772 & 0.6870 & 0.6859 & 0.7042 & 0.7151 & 0.7378 & 0.7379 & 0.7342 & 0.7562 \\
      \hline
      
      \multirow{4}{*}{\textbf{GAE}} 
      & Accuracy & 0.9740 & 0.9494 & 0.9215 & 0.8962 & 0.8723 & 0.8525 & 0.8271 & 0.7994 & 0.7781 & 0.7475 & 0.7198 & 0.7048 \\
      & FAR      & 0.0009 & 0.0016 & 0.0038 & 0.0077 & 0.0118 & 0.0156 & 0.0218 & 0.0303 & 0.0673 & 0.1061 & 0.0938 & 0.1812 \\
      & MDR      & 0.6050 & 0.5900 & 0.6017 & 0.5837 & 0.5680 & 0.5433 & 0.5400 & 0.5413 & 0.4794 & 0.4575 & 0.5005 & 0.4092 \\
      & $F_1$    & 0.5583 & 0.5744 & 0.5591 & 0.5722 & 0.5850 & 0.6075 & 0.6081 & 0.6039 & 0.6376 & 0.6416 & 0.6204 & 0.6668 \\
      \hline
      
      \multirow{4}{*}{\textbf{Transformer}} 
      & Accuracy & 0.9752 & 0.9508 & 0.9254 & 0.9000 & 0.8779 & 0.8633 & 0.8431 & 0.8196 & 0.8185 & 0.7990 & 0.7727 & 0.7638 \\
      & FAR      & \textbf{0.0000} & 0.0002 & 0.0017 & 0.0020 & 0.0058 & 0.0075 & 0.0106 & 0.0175 & 0.0257 & \textbf{0.0379} & \textbf{0.0427} & \textbf{0.0796} \\
      & MDR      & 0.5950 & 0.5875 & 0.5850 & 0.5900 & 0.5640 & 0.5242 & 0.5121 & 0.5062 & 0.4411 & 0.4295 & 0.4455 & 0.3929 \\
      & $F_1$    & 0.5765 & 0.5830 & 0.5818 & 0.5775 & 0.5981 & 0.6352 & 0.6446 & 0.6460 & 0.6979 & 0.7028 & 0.6910 & 0.7199 \\
      \hline
      
      \multirow{4}{*}{\textbf{CGCA-AL}} 
      & Accuracy & \textbf{0.9842} & \textbf{0.9669} & \textbf{0.9485} & \textbf{0.9356} & \textbf{0.9169} & \textbf{0.9085} & \textbf{0.8925} & \textbf{0.8869} & \textbf{0.8771} & \textbf{0.8667} & \textbf{0.8477} & \textbf{0.8442} \\
      & FAR      & \textbf{0.0000} & \textbf{0.0000} & \textbf{0.0012} & \textbf{0.0013} & \textbf{0.0029} & \textbf{0.0072} & \textbf{0.0076} & \textbf{0.0159} & \textbf{0.0243} & 0.0386 & 0.0462 & 0.0858 \\
      & MDR      & \textbf{0.3800} & \textbf{0.3975} & \textbf{0.4033} & \textbf{0.3800} & \textbf{0.3880} & \textbf{0.3442} & \textbf{0.3500} & \textbf{0.3075} & \textbf{0.2872} & \textbf{0.2660} & \textbf{0.2777} & \textbf{0.2258} \\
      & $F_1$    & \textbf{0.7654} & \textbf{0.7520} & \textbf{0.7435} & \textbf{0.7625} & \textbf{0.7542} & \textbf{0.7819} & \textbf{0.7791} & \textbf{0.8032} & \textbf{0.8131} & \textbf{0.8210} & \textbf{0.8130} & \textbf{0.8324} \\
      \hline
      
      & & 13 & 14 & 15 & 16 & 17 & 18 & 19 & 20 & 21 & 22 & 23 & 24 \\
      \hline
      
      \multirow{4}{*}{\textbf{GCN}} 
      & Accuracy & 0.7644 & 0.7529 & 0.7548 & 0.7698 & 0.7671 & 0.7946 & 0.8102 & 0.8323 & 0.8583 & 0.8860 & 0.9165 & 0.9510 \\
      & FAR      & 0.1914 & 0.2235 & 0.2439 & 0.2775 & 0.3121 & 0.3608 & 0.3860 & 0.4400 & 0.4317 & 0.4950 & 0.4950 & \textbf{0.0000} \\
      & MDR      & 0.2731 & 0.2639 & 0.2460 & 0.2066 & 0.2003 & 0.1536 & 0.1382 & 0.1133 & 0.1002 & 0.0793 & 0.0657 & 0.0490 \\
      & $F_1$    & 0.7697 & 0.7766 & 0.7935 & 0.8213 & 0.8295 & 0.8607 & 0.8779 & 0.8981 & 0.9175 & 0.9368 & 0.9554 & 0.9749 \\
      \hline
      
      \multirow{4}{*}{\textbf{GAT}} 
      & Accuracy & 0.7783 & 0.7762 & 0.7760 & 0.7771 & 0.7675 & 0.7865 & 0.7910 & 0.8065 & 0.8129 & 0.8369 & 0.8485 & 0.8675 \\
      & FAR      & 0.1341 & 0.1520 & 0.1633 & \textbf{0.1800} & \textbf{0.1936} & \textbf{0.1908} & 0.2180 & \textbf{0.2612} & \textbf{0.2400} & \textbf{0.2400} & \textbf{0.2800} & \textbf{0.0000} \\
      & MDR      & 0.2958 & 0.2750 & 0.2603 & 0.2444 & 0.2485 & 0.2211 & 0.2066 & 0.1800 & 0.1795 & 0.1561 & 0.1459 & 0.1325 \\
      & $F_1$    & 0.7749 & 0.7908 & 0.8050 & 0.8188 & 0.8208 & 0.8455 & 0.8574 & 0.8760 & 0.8847 & 0.9046 & 0.9153 & 0.9290 \\
      \hline
      
      \multirow{4}{*}{\textbf{GAE}} 
      & Accuracy & 0.7050 & 0.6969 & 0.7190 & 0.7398 & 0.7577 & 0.8015 & 0.8371 & 0.8575 & 0.8910 & 0.9271 & 0.9606 & \textbf{0.9923} \\
      & FAR      & 0.2405 & 0.2955 & 0.3722 & 0.4188 & 0.5050 & 0.5267 & 0.5570 & 0.6025 & 0.6233 & 0.6525 & 0.6750 & \textbf{0.0000} \\
      & MDR      & 0.3412 & 0.3086 & 0.2263 & 0.1809 & 0.1341 & 0.0892 & 0.0592 & \textbf{0.0505} & \textbf{0.0355} & \textbf{0.0202} & \textbf{0.0117} & \textbf{0.0077} \\
      & $F_1$    & 0.7076 & 0.7269 & 0.7748 & 0.8076 & 0.8351 & 0.8731 & 0.9014 & 0.9174 & 0.9394 & 0.9610 & 0.9796 & \textbf{0.9961} \\
      \hline
      
      \multirow{4}{*}{\textbf{Transformer}} 
      & Accuracy & 0.7742 & 0.7510 & 0.7677 & 0.7706 & 0.7817 & 0.8100 & 0.8273 & 0.8444 & 0.8715 & 0.9092 & 0.9452 & 0.9717 \\
      & FAR      & \textbf{0.0895} & \textbf{0.1205} & \textbf{0.1528} & 0.2050 & 0.2764 & 0.2958 & 0.3360 & 0.4138 & 0.4100 & 0.4700 & 0.5250 & \textbf{0.0000} \\
      & MDR      & 0.3412 & 0.3407 & 0.2800 & 0.2416 & 0.1944 & 0.1547 & 0.1297 & 0.1040 & 0.0883 & 0.0564 & 0.0343 & 0.0283 \\
      & $F_1$    & 0.7596 & 0.7555 & 0.7948 & 0.8151 & 0.8394 & 0.8697 & 0.8886 & 0.9056 & 0.9254 & 0.9501 & 0.9712 & 0.9856 \\
      \hline
      
      \multirow{4}{*}{\textbf{CGCA-AL}} 
      & Accuracy & \textbf{0.8390} & \textbf{0.8363} & \textbf{0.8354} & \textbf{0.8375} & \textbf{0.8408} & \textbf{0.8598} & \textbf{0.8715} & \textbf{0.8885} & \textbf{0.9129} & \textbf{0.9433} & \textbf{0.9654} & 0.9894 \\
      & FAR      & 0.1023 & 0.1240 & 0.1839 & 0.2394 & 0.2764 & 0.3192 & 0.3820 & 0.3937 & 0.4000 & 0.4225 & 0.4950 & \textbf{0.0000} \\
      & MDR      & \textbf{0.2108} & \textbf{0.1921} & \textbf{0.1530} & \textbf{0.1241} & \textbf{0.1109} & \textbf{0.0806} & \textbf{0.0618} & 0.0550 & 0.0424 & 0.0234 & 0.0146 & 0.0106 \\
      & $F_1$    & \textbf{0.8415} & \textbf{0.8520} & \textbf{0.8655} & \textbf{0.8779} & \textbf{0.8878} & \textbf{0.9077} & \textbf{0.9204} & \textbf{0.9339} & \textbf{0.9506} & \textbf{0.9693} & \textbf{0.9820} & 0.9947 \\
      \hline
      
      \end{tabular}}
      \end{table*}

    \begin{table*}[t]
      \centering
      \caption{Estimated error of LSI task for IEEE 118-bus case under different numbers of attacked lines ($|F|=1$--$24$).}
      \label{tab:lsicase118_1_24}
      \footnotesize
      \renewcommand{\arraystretch}{1.05}
      \setlength{\tabcolsep}{4pt}
      \resizebox{\textwidth}{!}{%
      \begin{tabular}{c|cccccccc}
      \hline
      \textbf{Algorithm} & \multicolumn{8}{c}{\textbf{Number of Attacked Lines} $|F|$} \\
      \hline
      & 1 & 2 & 3 & 4 & 5 & 6 & 7 & 8 \\
      \hline
      \textbf{FLD} &
      \mstd{0.3145}{0.4726} & \mstd{0.3591}{0.3844} & \mstd{0.3596}{0.3404} & \mstd{0.3929}{0.3215} &
      \mstd{0.3906}{0.2702} & \mstd{0.4029}{0.2375} & \mstd{0.3936}{0.2157} & \mstd{0.4180}{0.2065} \\
      \textbf{GCN+LSI} &
      \mstd{0.2646}{0.4240} & \mstd{0.2985}{0.3802} & \textbf{\mstd{0.2983}{0.3397}} & \mstd{0.3467}{0.3223} &
      \mstd{0.3426}{0.2877} & \mstd{0.3480}{0.2514} & \mstd{0.3409}{0.2283} & \mstd{0.3739}{0.2172} \\
      \textbf{GAT+LSI} &
      \mstd{0.2421}{0.4096} & \textbf{\mstd{0.2936}{0.3736}} & \mstd{0.3004}{0.3392} & \mstd{0.3466}{0.3222} &
      \mstd{0.3408}{0.2864} & \mstd{0.3483}{0.2510} & \mstd{0.3409}{0.2291} & \mstd{0.3738}{0.2180} \\
      \textbf{GAE+LSI} &
      \mstd{0.2454}{0.4166} & \mstd{0.3010}{0.3796} & \textbf{\mstd{0.2983}{0.3397}} & \mstd{0.3467}{0.3223} &
      \mstd{0.3423}{0.2880} & \mstd{0.3477}{0.2519} & \textbf{\mstd{0.3395}{0.2281}} & \mstd{0.3747}{0.2178} \\
      \textbf{Transformer+LSI} &
      \mstd{0.2611}{0.4200} & \mstd{0.2975}{0.3757} & \textbf{\mstd{0.2983}{0.3397}} & \mstd{0.3467}{0.3222} &
      \mstd{0.3409}{0.2865} & \mstd{0.3476}{0.2519} & \mstd{0.3414}{0.2293} & \mstd{0.3741}{0.2172} \\
      \textbf{CGCA-AL+LSI} &
      \textbf{\mstd{0.2407}{0.3969}} & \mstd{0.2980}{0.3760} & \mstd{\textbf{0.2983}}{0.3398} & \textbf{\mstd{0.3442}{0.3230}} &
      \textbf{\mstd{0.3379}{0.2850}} & \textbf{\mstd{0.3464}{0.2503}} & \mstd{0.3402}{0.2305} & \textbf{\mstd{0.3731}{0.2179}} \\
      \hline
      
      & 9 & 10 & 11 & 12 & 13 & 14 & 15 & 16 \\
      \hline
      \textbf{FLD} &
      \mstd{0.3903}{0.1972} & \mstd{0.4137}{0.1949} & \mstd{0.4294}{0.1451} & \mstd{0.4046}{0.1507} &
      \mstd{0.4139}{0.1438} & \mstd{0.3955}{0.1344} & \mstd{0.3850}{0.1372} & \mstd{0.3950}{0.1363} \\
      \textbf{GCN+LSI} &
      \mstd{0.3455}{0.2013} & \mstd{0.3626}{0.2028} & \mstd{0.3814}{0.1573} & \mstd{0.3460}{0.1647} &
      \mstd{0.3507}{0.1561} & \mstd{0.3348}{0.1394} & \mstd{0.3171}{0.1385} & \mstd{0.3246}{0.1412} \\
      \textbf{GAT+LSI} &
      \mstd{0.3435}{0.2033} & \mstd{0.3619}{0.2032} & \mstd{0.3814}{0.1558} & \mstd{0.3463}{0.1645} &
      \mstd{0.3519}{0.1567} & \mstd{0.3334}{0.1395} & \mstd{0.3145}{0.1390} & \mstd{0.3251}{0.1423} \\
      \textbf{GAE+LSI} &
      \mstd{0.3454}{0.2021} & \mstd{0.3642}{0.2020} & \mstd{0.3799}{0.1574} & \mstd{0.3471}{0.1639} &
      \mstd{0.3493}{0.1561} & \mstd{0.3348}{0.1388} & \mstd{0.3150}{0.1373} & \mstd{0.3212}{0.1448} \\
      \textbf{Transformer+LSI} &
      \textbf{\mstd{0.3417}{0.2049}} & \mstd{0.3640}{0.2034} & \mstd{0.3807}{0.1578} & \mstd{0.3467}{0.1662} &
      \mstd{0.3524}{0.1573} & \mstd{0.3317}{0.1412} & \mstd{0.3172}{0.1384} & \mstd{0.3250}{0.1428} \\
      \textbf{CGCA-AL+LSI} &
      \mstd{0.3451}{0.2008} & \textbf{\mstd{0.3585}{0.2061}} & \textbf{\mstd{0.3793}{0.1570}} & \textbf{\mstd{0.3417}{0.1639}} &
      \textbf{\mstd{0.3466}{0.1579}} & \textbf{\mstd{0.3298}{0.1411}} & \textbf{\mstd{0.3115}{0.1383}} & \textbf{\mstd{0.3184}{0.1457}} \\
      \hline
      
      & 17 & 18 & 19 & 20 & 21 & 22 & 23 & 24 \\
      \hline
      \textbf{FLD} &
      \mstd{0.3995}{0.1127} & \mstd{0.3822}{0.1076} & \mstd{0.3903}{0.1047} & \mstd{0.3936}{0.1004} &
      \mstd{0.3958}{0.0931} & \mstd{0.3821}{0.0891} & \mstd{0.3857}{0.0856} & \mstd{0.3897}{0.0749} \\
      \textbf{GCN+LSI} &
      \mstd{0.3253}{0.1195} & \mstd{0.2949}{0.1194} & \mstd{0.2993}{0.1136} & \mstd{0.2953}{0.1078} &
      \mstd{0.2865}{0.1032} & \mstd{0.2540}{0.0996} & \mstd{0.2481}{0.0957} & \mstd{0.2488}{0.0810} \\
      \textbf{GAT+LSI} &
      \mstd{0.3250}{0.1213} & \mstd{0.2926}{0.1214} & \mstd{0.2981}{0.1180} & \mstd{0.2954}{0.1096} &
      \mstd{0.2897}{0.1037} & \mstd{0.2551}{0.1006} & \mstd{0.2490}{0.0954} & \mstd{0.2466}{0.0867} \\
      \textbf{GAE+LSI} &
      \mstd{0.3239}{0.1205} & \mstd{0.2934}{0.1166} & \mstd{0.2986}{0.1144} & \mstd{0.2946}{0.1075} &
      \mstd{0.2867}{0.1022} & \mstd{0.2506}{0.0996} & \mstd{0.2473}{0.0960} & \mstd{0.2477}{0.0818} \\
      \textbf{Transformer+LSI} &
      \textbf{\mstd{0.3231}{0.1239}} & \mstd{0.2935}{0.1201} & \mstd{0.2991}{0.1167} & \mstd{0.2933}{0.1095} &
      \mstd{0.2918}{0.1042} & \mstd{0.2552}{0.1019} & \mstd{0.2485}{0.0977} & \mstd{0.2483}{0.0854} \\
      \textbf{CGCA-AL+LSI} &
      \mstd{0.3241}{0.1207} & \textbf{\mstd{0.2888}{0.1180}} & \textbf{\mstd{0.2929}{0.1119}} & \textbf{\mstd{0.2909}{0.1060}} &
      \textbf{\mstd{0.2821}{0.1037}} & \textbf{\mstd{0.2501}{0.1006}} & \textbf{\mstd{0.2471}{0.0964}} & \textbf{\mstd{0.2453}{0.0820}} \\
      \hline
      \end{tabular}}
      \end{table*}

    For performance comparison of LSI on IEEE 118-bus case, the results are given in Table~\ref{tab:lsicase118_1_24}. All learning-based LSI algorithms achieves significantly lower average estimated error while only having a little higher standard deviation when $|F| \ge 3$. Specifically, CGCA-AL+LSI always achieves the best or near-best performance on estimated error in most cases, demonstrating strong attack diagnosis capability in large-scale networks. Quantitatively, for moderate attack scenarios ($|\mathcal{F}|=4$--12), CGCA-AL+LSI reduces the estimated error by approximately $1\%$--$2\%$ compared with representative learning-based baselines (GCN/GAT/Transformer+LSI), while for dense attacks ($|\mathcal{F}|\geq 17$), the advantage remains stable, indicating good scalability. Moreover, compared with the model-based FLD approach, CGCA-AL+LSI consistently yields substantially lower estimation errors (often exceeding $30\%$ reduction), highlighting its superior robustness and accuracy under complex and large-scale attack conditions.

    In general, these two sets of numerical simulation results demonstrate the excellent and robust performance of the proposed attack diagnosis scheme, especially in scenarios with complex attacks.

    \section{Conclusion}
    In this study, a multi-scale attention-enhanced attack diagnosis framework for power grids under PCPA was presented. The proposed framework consisted of two stages: support prior extraction using the CGCA-AL algorithm and learning-based LSI (formulated via MMIP). The CGCA-AL algorithm extracted spatial and long-range dependency patterns of power grids under PCPA by integrating GAT and CNN. The high-dimensional features of attacked nodes were further enhanced through a MHCA mechanism, which contrasted the original node features with learned high-level representations of observed nodes. Based on these enhanced features, representative embeddings of attacked transmission lines were constructed and used to generate informative support priors for the subsequent learning-based LSI stage. By solving the resulting learning-based LSI problem, accurate attack diagnosis under PCPA was achieved. Simulation results on the IEEE 30-bus and IEEE 118-bus cases demonstrated the effectiveness, robustness, and scalability of the proposed attack diagnosis framework for power grids subject to complex PCPA scenarios.

    There are several possible directions for further studies. One is to consider PCPA with more sophisticated, arguably more hostile cyber attacks, e.g., a carefully planned combination of DoS and FDI attacks. Another one, as mentioned earlier, is to design effective algorithms where Assumption 2 does not hold. Yet another development we would like to make is to investigate the effects and develop solutions when there exist strong noises in measurement data.

    \section*{Acknowledgment}
    Parts of this article have been grammatically
    revised using ChatGPT \cite{openai2023chatgpt} to improve readability. The code accompanying this paper is publicly available at the \url{https://github.com/ryan-ntu/Fault-Diagnosis_PCPA}.

    \bibliographystyle{IEEEtran}
    \bibliography{IEEE_abb, mybib}

\newpage
\clearpage

\appendix
\subsection{Proofs}
\subsubsection{The proof of Lemma \ref{IdentifyFBus}}
\label{proof:lemma1}
\begin{proof}
    From \eqref{addimitancematrixchange} and [Lemma 1, \cite{soltan2015joint}],
    it is easy to have the following equation:
    \begin{equation}
        \begin{aligned}
             & A\vec{\theta}- A'\vec{\theta}' = \vec{\Delta}                                                                                           \\
             & \Rightarrow \operatorname{supp}(A(\vec{\theta}- \vec{\theta}') - \vec{\Delta}) \subseteq \bigcup_{e_i \in F}e_{i}\subseteq{\cal V}_{H}.
        \end{aligned}
    \end{equation}
    For more details, please refer to \cite{soltan2015joint}.
\end{proof}

\subsubsection{The proof of Lemma \ref{le: 2}}
\label{proof:lemma2}
\begin{proof}
    From Lemma \ref{IdentifyFBus}, we have
    \begin{equation}
        \label{BarH}A_{\bar{H}|H}\vec{\theta}_{H}'= A_{\bar{H}|H}\vec{\theta}
        _{H}+A_{\bar{H}|\bar {H}}(\vec{\theta}_{\bar H}-\vec{\theta}_{\bar H}
        ') - \vec{\Delta}_{\bar H}.
    \end{equation}
    Since the right-hand side of \eqref{BarH} is known, the reconstruction of $\vec
    {\theta}_{H}'$ is only dependent on the column rank of $A_{\bar {H}|H}$.
    Based on Assumption \ref{assum:1}, the sufficient condition for the existence
    of the unique solution of $\vec{\theta}_{H}'$ thus is that $A_{\bar{H}|H}$
    has a full column rank.
\end{proof}

\subsubsection{The proof of Lemma \ref{le: deltarecovery}}
\label{proof:lemma3}
\begin{proof}
    According to Assumption \ref{assum:2}, for any bus $u \in{\cal V}_{H}$, there
    exists an adjacent bus $v \in{\cal V}_{\bar H}$ such that $e = (u, v) \in
    \cal{E}$. Since $v$ is outside the attacked area and its power injection
    change is known, we can compute the proportion $p'_{v}/ p_{v}$. Applying
    \eqref{eq: powerrecovery}, we can then determine the proportional change
    in power injection at bus $u$. Consequently, $\vec{p}'_{H}$ and $\vec{\Delta}
    '_{H}$ can be reconstructed.
\end{proof}

\subsection{Explanation of Assumption~\ref{assum:2}}
\label{appen:assump2}
These primary extreme scenarios are predominantly categorized into two sets, with special reactance values and unconnected buses, respectively. In the first set of scenarios, it occurs with a particular set of reactance values of ${\cal E}\left[{\cal V}_{H},{\cal V}_{\bar H}\right]$ for which the column rank of $A_{\bar H|H}$ is not complete though
${\cal V}_{H}$ is covered by ${\cal E}\left[{\cal V}_{H},{\cal V}_{\bar H}\right]$. In reality, this is basically impossible because it requires at least two buses $v_{i}$ and $v_{j}$ within $H$ that for any $u \in{\cal{V}}_{\bar H}$, the admittance values of transmission lines ${v_i, u}$ and ${v_j, u}$ are equal. Considering the complexity of power grids, such a set is a measure zero set in real space (see \cite{soltan2018power}). The second set of scenarios occurs only when some buses in ${\cal V}_{H}$ do not have a direct
connection to $\bar H$. In this context, the consequence of this lack of connectivity is that $A_{{\bar H}|H}$ has an all-zero column. MTD may provide a good solution to infer lost information in the first set of scenarios, but it is ineffective in handling the second set of scenarios, as MTD can only change the admittance of transmission lines, not the connectivity of power grids. It means that MTD may need multiple trials to achieve the situation awareness of the attacked area with great risks of cascading failures. Therefore, MTD may fails for second cases. The remedy, which may involve installing secured PMUs on carefully selected buses (see \cite{pei2020pmu}), requires further investigations. 

\subsection{DBGS Algorithm}
The attacked area of PCPA within power grids is contructed based on DBGS algorithm, as shown in Algorithm~\ref{alg:dbgs}.
\begin{algorithm}[t]
    \caption{Degree-Based Greedy Search (DBGS) for Constructing PCPA Attacked Area}
    \label{alg:dbgs}
    \begin{algorithmic}[1]
        \STATE \textbf{Input:} Power grid graph $\mathcal{G}=(\mathcal{V},\mathcal{E})$, target size $|\mathcal{V}_H|$, Assumptions~\ref{assum:1}--\ref{assum:2}.
        \STATE \textbf{Output:} Attacked subgraph $(\mathcal{V}_H,\mathcal{E}_H)$.
        \STATE Randomly select a seed bus $v_{0}\in\mathcal{V}$.
        \STATE Initialize $\mathcal{V}_H\gets\{v_{0}\}$.
        \WHILE{$|\mathcal{V}_H|<|\mathcal{V}_H|_{\text{target}}$}
            \STATE Identify candidate neighbor set 
            \[
            \mathcal{N}(\mathcal{V}_H)=\{v\in\mathcal{V}\setminus\mathcal{V}_H \mid \exists (u,v)\in\mathcal{E},\, u\in\mathcal{V}_H\}.
            \]
            \FOR{$v\in\mathcal{N}(\mathcal{V}_H)$}
                \STATE Compute external degree 
                \[
                d_{\mathrm{ext}}(v)
                =\left|\{(v,u)\in\mathcal{E}\mid v\in\mathcal{V}\setminus\mathcal{V}_H\}\right|.
                \]
            \ENDFOR
            \STATE Select $v^{\star}=\arg\max_{v\in\mathcal{N}(\mathcal{V}_H)} d_{\mathrm{ext}}(v)$.
            \STATE Update $\mathcal{V}_H\gets\mathcal{V}_H\cup\{v^{\star}\}$.
        \ENDWHILE
        \STATE Construct attacked edge set
        \[
        \mathcal{E}_H=\{(u,v)\in\mathcal{E}\mid u\in\mathcal{V}_H,\, v\in\mathcal{V}_H\}.
        \]
        \RETURN $(\mathcal{V}_H,\mathcal{E}_H)$.
    \end{algorithmic}
\end{algorithm}

\subsection{Performance metrics and training platform}
\label{training_details}
    \subsubsection{Metrics}
    To evaluate the performance of our CGCA-AL algorithm, we introduce
    the accuracy, false alarm rate (FAR), missed detection rate (MDR) and $\text{F}
    _{1}$ score. Their definitions are as follows:
    \begin{align}
        \label{} & \text{Accuracy}= \frac{\text{TP}+ \text{TN}}{\text{TP}+\text{FP} +\text{TN}+\text{FN}}, \\
                 & \text{FAR}= \frac{\text{FP}}{\text{FP}+\text{TN}},                                      \\
                 & \text{MDR}= \frac{\text{FN}}{\text{TP}+\text{FN}},                                      \\
                 & \text{F}_{1}= \frac{2*\text{TP}}{2 \text{TP}+ \text{FP} + \text{FN}},
    \end{align}
    where TP, FP, TN, and FN denote true positives, false positives, true
    negatives, and false negatives, respectively. To evaluate the performance of
    the LSI algorithm, we introduce the normalized estimated error term
    \begin{equation}
        \text{error}= \frac{\|\vec{x}^*_{H}- \vec{x}_H\|}{\|\vec{x}_{H}\|},
    \end{equation}
    where $\vec{x}^*_{H}$ denotes the ground truth of physical attack
    vector and $\vec{x}_H$ is the estimation of physical attack vector provided by
    the LSI algorithm.

    \subsubsection{Training Platform}
    All algorithms run on AMD Ryzen Threadripper PRO 3955WX 16-Cores 3.9GHz CPU with
    $4$ NVIDIA GeForce RTX 3090 GPUs. The implementation of GAT and GCN model is
    based on PyTorch in Python.

    \subsubsection{Baselines}
    In this simulation, we introduce several baseline methods to benchmark the performance of the proposed CGCA-AL algorithm and the corresponding LSI-based diagnosis framework. The comparison is conducted in two stages: AL and LSI. For the AL stage, the proposed CGCA-AL is compared with four baselines: (1) a \textbf{GCN}-based method reported in \cite{chen2020fault, boyaci2022joint}; (2) a \textbf{GAT}-based method introduced in \cite{zhang2022graph}; (3) a \textbf{graph autoencoder (GAE)} approach from \cite{fahim2024graph}; and (4) a \textbf{Transformer}-based model proposed in \cite{thomas2023cnn}. To ensure a fair comparison, the GAE and Transformer models are slightly adapted to the attack localization task by aligning their input and output formats, while their core architectures are preserved. For the LSI stage, five baselines are considered, including four LSI algorithms corresponding to the above AL baselines, as well as the FLD algorithm proposed in \cite{huang2022link}.
    \subsubsection{Parameter setting}
    The \textbf{CGCA-AL} algorithm consists of three GAT--CNN layers with channels $(256,256,256)$ and attention heads $(4,4,4)$, followed by a MHCA layer with 256 channels and 4 heads, and a dense prediction head with channels $(256,128,128)$. The \textbf{GCN} baseline includes three GCN layers with channels $(256,256,256)$ and the same dense head. The \textbf{GAT} baseline employs three GAT layers with channels $(256,256,256)$ and heads $(4,4,4)$, followed by the identical dense head. The \textbf{GAE} baseline consists of three GCN layers with channels $(256,256,256)$, a self-attention layer with 256 channels and 4 heads, and the same dense head. The \textbf{CNN--Transformer} baseline comprises a node tokenizer with channels $(128,256)$, positional encoding with 256 channels, a two-layer Transformer encoder with model dimension 256 and 4 attention heads, as well as the same dense head.

    For all algorithms, the same set of hyperparameters is adopted to ensure a fair comparison. Specifically, the batch size is set to 128 and 256 for the IEEE 30-bus and IEEE 118-bus cases, respectively. The maximum number of training epochs is 300, with an early-stopping patience of 20 epochs. The learning rate and weight decay are both set to $10^{-4}$. The Adam optimizer is used with the binary cross-entropy (BCE) loss function.

\vfill

\end{document}